\documentstyle[prl,aps]{revtex}
\begin{document}
\title{A hierarchical scheme for cooperativity and folding in proteins}
\author{Alex Hansen, 
\footnote{Permanent address: Department of Physics, 
Norwegian University of Science and Technology, 
NTNU, N--7034 Trondheim, Norway}
Mogens H. Jensen, Kim Sneppen and Giovanni Zocchi}
\address{Niels Bohr Institute and NORDITA, Blegdamsvej 17, DK-2100 {\O}, 
Denmark}
\date{\today}
\maketitle
\begin{abstract}
We propose a protein model based on a hierarchy of constraints 
that force the protein to follow certain pathways when changing 
conformation.  The model exhibits a first order phase transition,
cooperativity and is exactly solvable.  It also shows an 
unexpected symmetry between folding and unfolding pathways
as suggested by a recent experiment.
\end{abstract}

Proteins function by switching between different well-defined conformations 
in the native state. In order to reach the native state the 
protein has to fold.  The number of possible conformations is huge, and
the existence of folding pathways guiding the proteins between the
relevant conformations seems necessary \cite{L68}.  
The existence of pathways has been experimentally established 
\cite{C74,C85,F93},
but the mechanisms behind these are still to be understood.  
The free energy difference
stabilizing the folded state of a protein, is of the order 20 
$k_B T_{room}$ for a typical domain consisting of about 100 amino acids, 
meaning that the binding energy per amino acid is only of order 
of $k_B T_{room}$ \cite{P92}. This is a surprisingly small value.  
The mechanism responsible for stability at very little cost is 
not known, but has been named ``cooperativity" \cite{P92}.

In this letter we propose to model protein folding
by a parametrization of the folding pathway.
Cooperativity appears in the theory through assuming that only 
interlocked degrees of freedom are active in the energy 
window that separate the folded and unfolded states. This also
leads to a first order folding transition.

Furthermore, the model exhibits an interesting dynamical symmetry when changing 
conformation. One would expect it to follow the folding pathway in 
opposite direction when it unfolds.  However, under certain conditions it 
moves in the same direction along the folding pathway both when folding and 
unfolding.  We call this the ``mirror effect."  This feature of the model
is in agreement with recent measurements \cite{Z96} which suggest this mirror 
symmetry in the conformational change of albumin at low pH.

The amino acids constituting a protein may rotate relative to each other, 
thus inducing conformational changes in the protein.  They furthermore 
interact. Through changing bond angles, amino acids far apart on the protein 
may be brought close to each other to form bonds that stabilize a given 
conformation.  

Simple models for protein folding which encompass guiding 
\cite{Z95} and cooperativity \cite{D93} have been
proposed. More complex models based on spin glass hamiltonians \cite{A85}
need to introduce ``folding funnels" 
in the energy landscape in order to guide the slow dynamics
of a frustrated system \cite{Bryng95,Dill95}.
The existence of a pathway 
implies that the bond angles change in a certain order, 
i.e.\ there is a hierarchy among the bond angles themselves. 
In order to clarify the significance of such a hierarchy, 
we introduce a model that simultaneously displays a folding pathway,
cooperativity and a first order phase transition,
thus reproducing three important characteristics of real proteins.
The model may be visualized as in Fig.\ \ref{fig:0}.  
The variables are effective folding angles $\varphi_1$ 
to $\varphi_N$, each associated with matching a given sub part 
of the protein into its position in the ground state (the template).
The key point of the model is that a correct fold at level $i$ 
gains binding energy when all lower levels are correctly folded. 
Our angle variables are not to be thought of as the $\phi,\psi$ 
angles of the polypeptide backbone, but rather as coarse grained 
variables (i.e.\ averaged over a small number of 
monomers).  An intuitive picture would be the angle between two $\alpha$ 
helices connected by a loop. Our model corresponds to the interaction 
Hamiltonian
\begin{equation}
\label{nat1}
H=-a_1\ \varphi_1-a_2\ 
\varphi_1\varphi_2-a_3\ \varphi_1\varphi_2\varphi_3-\cdots
-a_N\ \varphi_1\varphi_2\cdots\varphi_N\;,
\end{equation}
where $a_1$ to $a_N$ are coupling constants. 

In order to present explicit calculations we scale all angle variables 
$\varphi_i$ to be between $0$ and $1$ 
and set all coefficients $a_i$ equal to 1.
Furthermore, we assume for the moment that angle variables only take discrete 
values, 0 and 1.  The Hamiltonian then represents the number of folded 
variables ($\varphi_i=1$) counted from $i=1$ until the first $i=n_f+1$ where 
$\varphi_i=0$, and can simply be written
\begin{equation}
\label{eq2}
H= - n_f\;.
\end{equation}

To explore the thermodynamic properties of the model, we 
calculate its partition function by summing over degenerate states,
\begin{equation}
\label{eq3}
Z(T,N) \;=\; \sum_{n_f=0}^{N-1} 2^{N-1-n_f} e^{n_f/T}+ e^{N/T}\;=\;
2^{N-1} \; \frac{ \exp( N(1/T- \ln 2 ) ) -1 }
{ \exp(1/T- \ln 2 ) - 1 }
\; +\; e^{N/T}\;,
\end{equation}
where $T$ is temperature measured in units of the coefficients $a_i=1$.
In the limit $N\rightarrow \infty$, the partition function has a discontinuous 
derivative for $T=T_c=1/\ln 2$. This means that the system undergoes a
first order phase transition at this temperature.  At temperatures 
$T<T_c$ the binding energy of the folds dominates and the system freezes into
the folded state with $\langle n_f\rangle \approx N$, whereas for $T>T_c$ 
the winning term will be the entropy that is gained from each folded variable, 
yielding $\langle n_f\rangle \approx 0$, corresponding to an unfolded, 
molten state.  
Experimentally, it is known that small proteins typically undergo
a first order transition, while for large, multidomain proteins the
situation can be more complex \cite{P92}. 

In addition to a phase transition, the model exhibits cooperativity,
manifested through the fact that below $T_c$, say for $T=1$,
the chain is folded even though the binding energy per link is only 
$kT$, which would lead to a disordered (unfolded) state in the
absence of cooperativity. In the present model, at $T< T_c$, 
nearly all states will match the template, with fluctuations 
only at the bottom of the hierarchy. This matches the findings from
real proteins, where cooperativity implies that most degrees 
of freedom are locked by others.

During a slow ``adiabatic'' change of temperature through the phase 
transition, the folding/unfolding follow specific pathways.  Under cooling, 
the first level in the hierarchy must fold first, $\varphi_1=0 \to 1$ in order 
for the energy to change at all. This is due to the nature of the Hamiltonian 
where $\varphi_1$ is present in all the terms, see (\ref{nat1}). After the 
first, the next level in the hierarchy, $\varphi_2$, must fold in order to 
change the energy further and so on. Under heating, the unfolding will take 
place from bottom and up, i.e. first $\varphi_N= 1 \to 0$ leading finally to 
$\varphi_1=1 \to 0$. This is due to the fact that the smallest increase of 
energy is given by the unfolding of the deepest levels in the hierarchy.

Under non-adiabatic changes the unfolding is however different.  When $T<T_c$, 
the free energy is 
\begin{equation}
\label{fh}
F_E(T,N)=-N\;,
\end{equation}
as it is dominated by the binding energy.  For $T>T_c$, the free energy is  
\begin{equation}
\label{fe}
F_S(T,N)=-T N \ln 2\;,
\end{equation}
which results from entropy now dominating. We denote $F_E$ the energy
branch and $F_S$ the entropy branch of the free energy.  We may envision
these extended beyond the critical temperature $T_c$ as shown in Fig.\
\ref{fig:2}.  
We now perform a quench of the system, i.e., rapidly change its temperature.  
Starting in equilibrium at a temperature $T < T_c$, marked ``1"
in Fig.\ \ref{fig:2}, and rapidly heat the system to $T > T_c$, it moves to 
``2" in the figure, following the energy branch of the free energy.  The 
system then thermalizes by moving to ``3", after which the system ends up on 
the entropy branch.  Starting with the system in equilibrium at a temperature 
$T>T_c$, point ``4" in the figure, a rapid quench to a low temperature brings 
the system to ``5" along the entropy branch.  The system then moves from ``5" 
to ``6" to reach the energy branch which at this temperature is the lowest 
free energy.

The question is now how the angle variables react to these two types of quench.
When the system is rapidly cooled, the angle variables have to fold in the 
order 1,2,3,... $N$ --- otherwise, no change in energy occurs.  This is the 
same order as the system follows during adiabatic cooling.  However, when 
going from low to high temperature the situation is different.  When the system
moves from ``2" to ``3" in Fig.\ \ref{fig:2}, the unfolding rate, $\Gamma_i$, 
for each $\varphi_i$, is dominated by the size of the phase space of the final
state, 
\begin{equation}
\label{rate}
\Gamma_i\; \propto e^{i(1/T-\ln 2)} \;.
\end{equation}
Therefore the rate increases with decreasing $i$, i.e., the rate of 
change is largest for $\varphi_1$, second largest for $\varphi_2$ etc.  This 
is the same order as when the system was rapidly cooled and the opposite to 
the order found in adiabatic melting.  We call this reversal of direction 
``mirror effect." So far as we have regarded the variables $\varphi_i$ as 
binary, as soon as the first angle has unfolded, the others will unfold 
in no particular order, see Fig.\ \ref{fig:0}. With continuous variables this 
degeneracy of the model is lifted.

In a recent experiment, an indication of the existence of this 
``mirror effect" was found \cite{Z96}. In the experiment, the 
conformational change induced in albumin by lowering pH is
studied on samples consisting of very few (1-100) molecules. 
The technique is described in detail in \cite{Z96}; briefly, the 
molecules are sandwiched between a flat plate and a micron size 
sphere, and the movement of the sphere relative to the plate, 
measured to sub nm resolution with an evanescent wave technique, 
is used to reveal changes in shape (e.g. swelling or contracting) 
of the molecules in response to a fast pH change. This corresponds 
to quenching the proteins into the new state. 

Albumin is a large protein ($\sim$ 600 aminoacids), consisting of 6 
domains. It is believed that the conformational change at low pH 
consists essentially in a rearrangement of these domains, so 
that the shape of the molecule changes (becoming about 60 \% 
longer). However, this is not a complete unfolding: the molecule 
is still compact, and elements of secondary structure persist. 
In Fig.\ \ref{fig:1} we present some measurements from the experiment 
\cite{Z96}. As described in \cite{Z96}, the conformational 
change is seen to proceed through a series of $\sim$ 5 steps in time, 
suggesting a sequential rearrangement of the domains. Thus for this 
process, a well defined pathway exists. 

The experiment further suggests that the reverse conformational change 
(obtained by restoring neutral pH conditions) proceeds through the same 
sequence (not the time reversed sequence) of steps. We observe that the steps 
occur at similar positions, supporting a mirror imaging of the retracing
pathway.  In support of this observation, we project in Fig.\ \ref{fig:3} the 
trajectory on the distance axis, to show a histogram of the time spend at 
each distance value, i.e.\ such that flat regions (waiting times) will be 
represented by peaks in the histogram.  We can then average several histograms.
The comparison of the two symmetries, i.e. the ``mirror effect'' (b) and
simple time reversal (c), supports that the proteins under these conditions 
show the ``mirror effect.''

In conclusion, we believe to have introduced the first model
which simultaneously reproduces three key features of real protein folding,
namely a folding pathway, cooperativity and a first order folding transition.
In addition the model raises some interesting symmetry 
discussions (i.e. the ``mirror effect'').

A.H.\ thanks Nordita for funding his stay in Copenhagen during which
this work was done. Also we like to thank R.\ Donangelo and
K.\ B.\ Lauritsen  for discussions.


\begin{figure}
\caption[x]{
The figure describes the refolding of a connected structure, such as
a partly unfolded protein, into a well defined template.
It illustrates schematically
how one by successive foldings of angle variables
can match the edges 1,2,3.. to the template edges 1',2',3'...
We suggest to copy the figure and do  the folding.
One will see that it is not possible to match say edge 2 to template 2',
without first matching 1 to 1'. In general to match
$j$ with $j'$ demands correct matchings for all $i<j$.
This defines a unique pathway from the unfolded to the folded
structure, in a way that can be quantified by the 
Hamiltonian in Eq.\ (\ref{nat1}).
\label{fig:0}
}
\end{figure}
\begin{figure}
\caption[x]{
The free energy per variable. The
full line show the equilibrium free energy as a function of temperature.
The dashed lines show the entropic and energetic branches of the
free energy, respectively (Eqs.\ (\ref{fe}) and (\ref{fh})).
Path $1\to 2$ shows a rapid heating of the system.
It then unfolds following path $2\to 3$.  Path $4\to 5$ shows a
rapid cooling of the system followed by the folding following
path $5\to 6$.
\label{fig:2}
}
\end{figure}
\begin{figure}
\caption[x]{Curves from the experiment \cite{Z96} showing steps in the 
dynamics of the forward (left curve) and backward (right curve) conformational 
change.  The ordinate is the height of the sphere above the plate, $h$, 
in nm, and the abscissa is the time in seconds with arbitrary zero point. 
For the right curve, only the distance axis has been reversed. 
The two curves have been placed next to each other for ease of 
comparison.  The dashed lines help identify the steps. The correspondence 
between the steps in the two curves is the ``mirror effect."
\label{fig:1}
}
\end{figure}
\begin{figure}
\caption[x]{
Histograms of the time spent at each distance value, constructed from 
projecting curves of the type shown in Fig.\ \ref{fig:1} on the distance axis. 
The waiting time regions in the original curves then give rise to peaks 
in the histogram and the distance between two peaks is a step size. We 
then average several histograms. 
(a) is an average over histograms corresponding to three unfolding curves. 
The large peaks at ~8 and ~11 nm correspond to the two baselines
in Fig.\ \ref{fig:1}. The four peaks in between correspond to the four
waiting times shown by the dotted lines on Fig.\ \ref{fig:1}, left curve.
(b) is an average over histograms corresponding 
to three folding curves transformed by 
reversing the distance axis (i.e. according to the ``mirror effect''). Here 
there are only three
peaks corresponding to the three steps in Fig.\ \ref{fig:1},
right curve. As can be seen by comparing histograms
(a) and (b) the missing step is either covered by the peak of the
base line or simply does not occur in refolding. 
(c) presents the same histogram as in (b) but with simple time reversal,
i.e. no reversal of the diameter axis.
We see that (b) leads to better overlap with (a) than (c),
indicating that the proteins under these conditions follow the 
``mirror effect."
\label{fig:3}
}
\end{figure}

\begin{thebibliography}{10}
\bibitem{L68} C.\ Levinthal, J.\ de Chimie Physique et du 
Physico-Chimie Biologique {\bf 65} (1968) 44.
\bibitem{C74} T.\ E.\ Creighton, J.\ Mol.\ Biol.\ {\bf 87} (1974) 603.
\bibitem{C85} T.\ E.\ Creighton, J.\ Phys.\ Chem. {\bf 89} (1985) 2452.
\bibitem{F93} A.\ R.\ Fersht, FEBS {\bf 325} (1993) 5.
\bibitem{P92} P.\ L.\ Privalov, {\sl Protein folding,\/} ed.\ by
T.\ E.\ Creighton, 83 (Freeman, New York, 1992).
\bibitem{Z96} G.\ Zocchi, Proc.\ Natl.\ Acad.\ Sci., in print (1997).
\bibitem{Z95}
R.\ Zwanzig, Proc.\ Natl.\ Acad.\ Sci.\ {\bf 92} (1995) 9801.
\bibitem{D93} K.\ A. Dill, K.\ M. Fiebig, H.\ S.\ Chan, 
Proc.\ Natl.\  Acad.\ Sci.\ {\bf 90} (1993) 1942.
\bibitem{A85} A. Ansari, J. Berendzen, S.\ F. Browne,
H. Frauenfelder, I.\ T. Iben, T.\ B. Sauke, E. \ Shyamsunder and 
R.\ D.\ Young, Proc.\ Natl.\ Acad.\ Sci., {\bf 82} (1985) 5000.
\bibitem{Bryng95}
J.D.\ Bryngelson, J.N.\ Onuchic, N.D.\ Socci, P.G.\ Wolynes
Proteins: Struc.\ Funct.\ Genet.\ {\bf 21} (1995) 167.
\bibitem{Dill95} K.A.\ Dill, S.\ Bromberg, K.\ Yue, K.M.\ Fiebig, D.P.\
Yee, P.D.\ Thomas and H.S.\ Chan, Protein Science {\bf 4} (1995) 561.
\end{thebibliography}
\end{document}